# THE FCC-HCP TRANSITION IN SINGLE-COMPONENT AND BINARY VAN DER WAALS CLUSTERS FROM ELECTRON DIFFRACTION DATA

O.G. Danylchenko, S.I. Kovalenko, O.P. Konotop, V.N. Samovarov

*B. Verkin Institute for Low Temperature Physics and Engineering,*

*47 Lenin Ave., 61103 Kharkov, Ukraine*

*e-mail: konotop@ilt.kharkov.ua*

An electron diffraction study of nucleation and growth of the hcp phase in large ($2 \cdot 10^3$ – $1 \cdot 10^5$ at./cl.) pure Ar and mixed Ar-Kr clusters formed by supersonic jet expansion is carried out. The threshold cluster size corresponding to the appearance of an hcp phase in addition to the fcc structure is determined. It is found that the relative volume of the hcp phase in clusters increases with their size. In very large ($\delta \geq 150$ Å) aggregations, the percentage of hcp phase reaches its maximum and does not change with further cluster growth. The hcp phase relative volume in mixed clusters is almost double that in pure Ar clusters of the same size. A correlation between the hcp phase relative volume and the number of defective planes contained in the fcc matrix, which are the nuclei of the hcp phase, is established. A mechanism of nucleation and growth of the hcp phase in rare gas clusters is proposed.

Under normal conditions, all rare gases except helium have a face-centered cubic lattice, although theoretical calculations of the crystal potential energy in the two-body Lennard-Jones model show a 0.01% advantage of the hexagonal close packing. To realize a stable hcp structure in bulk samples ultrahigh pressures are applied to rare gases [1, 2]. Multiphase hcp-fcc [3] or pure hcp (for Ne) [4] structures are also prepared by annealing thin films deposited on a cold (2-10 K) substrate.

The fcc-hcp dilemma is an important issue in the studies of free single-component and binary rare gas clusters produced by adiabatic expansion of supersonic jets. The hcp structure was observed for the first time in large free Ar clusters by using electron diffraction technique [5]. Subsequently, the dependence of the fcc→hcp+fcc transition on size of these clusters was studied [6]. The hcp maxima were also found to appear in addition to the fcc ones in electron diffraction patterns from large ($\bar{N} \geq 2\cdot10^4$ at./cl.) heterogeneous Ar-Kr and Kr-Xe clusters [7]. But the fcc-hcp transition itself in binary van der Waals clusters has not been investigated yet. The purpose of this study is to compare the processes of nucleation and growth of the hcp structure in single-component Ar and binary Ar-Kr clusters. Additionally, it was also planned to obtain data on size dependence of the number of stacking faults in these clusters.

The experimental method for producing homogeneous and heterogeneous rare gas clusters as well as the technique used to study their structure was described in detail in Refs. [5, 8]. Cluster size was changed by varying the gas (gas mixture) backing pressure $P_0$ from 1 to 8 bar and the nozzle inlet temperature $T_0$ from 175 to 125 K. The component concentration in the heterogeneous clusters was kept at 50 ± 15 at. %.

The mean linear characteristic size of the fcc clusters as well as the size of fcc domains in two-phase clusters $\delta$ was determined from the broadening of electron diffraction peaks using the Scherrer relation and taking into account the presence of stacking faults (SF). The SF density $\alpha$, which is equal to $n_{sf}/n$ (here $n_{sf}$ is the number of defective planes and $n$ is the total number of planes), was calculated from the shift of the fcc (111) peak from its position in the defect-free crystal [9]. The total number of planes in the fcc crystal is $n = 3\delta/d_{111}$, here $d_{111}$ is the distance between the close-packed layers. If we determine $\alpha$ and calculate $\delta$, we can find the value $n_{sf}$. The average number $\bar{N}$ of atoms in clusters was found from the ratio of cluster volume to volume per 1 atom. The cluster volume was calculated in the spherical approximation considering that $\delta$ is the diameter of the spheres. The volume of the two-phase clusters is $V = V_{fcc} + V_{hcp}$, thus the characteristic linear size is $\delta = \sqrt[3]{\delta_{fcc}^3 + \delta_{hcp}^3}$.

The presence of the fcc diffraction peaks distinct from the hcp (100), (101), and (103) maxima indicates the existence of an hcp structure in the clusters. Subsequently determining the ratio of the integrated intensity of the hcp (101) maxima to fcc (200) one $I_{(101)}/I_{(200)}$ for clusters of different sizes, we calculated the $V_{hcp}/V_{fcc}$ value. For this purpose we used the well-known formula for the intensity of the diffracted electrons [10] abbreviated by cancellation of factors common to both phases:

$$\frac{I_{101}}{I_{200}} = \left|\frac{\Phi_{101}}{\Phi_{200}}\right|^2 \left(\frac{\Omega_{fcc}}{\Omega_{hcp}}\right)^2 \frac{P_{101} d_{101}^2}{P_{200} d_{200}^2} \frac{V_{hcp}}{V_{fcc}} \approx 1.2 \frac{V_{hcp}}{V_{fcc}}, \qquad (1)$$

here $\Phi_{hkl}$ are the structure factors for the (hkl) family of planes; $\Omega_{fcc}$ and $\Omega_{hcp}$ are the unit cell volumes of fcc and hcp phases, respectively; $P_{hkl}$ are the multiplicity factors; $d_{hkl}$ are the interplanar distances.

The above shown technique allowed us to quantitatively describe the fcc → hcp transition as a function of size of Ar and Ar-Kr clusters (Fig. 1). As can be seen in Fig. 1, only fcc structure is realized in small clusters ($\delta \leq 80$ Å). Weak maxima of the hcp phase were found in clusters with $\delta \approx 95$ Å ($\bar{N} \approx 11000$ at./cl.). The subsequent increase in the cluster size was accompanied by a rapid growth of the ratio $V_{hcp}/V_{fcc}$. Furthermore, the hcp phase relative volume in heterogeneous clusters is almost twice as great as its content in homogeneous clusters of the same size. Further increase in cluster size (δ ≥ 150 Å) caused a significant slowdown of the hcp phase accumulation rate.

To understand the mechanism of formation of the hcp phase in clusters we found the number of the defective planes $n_{sf}$ in the fcc lattice. Areas of the crystal lattice near such planes are the nuclei of the hcp phase in the fcc matrix. Fig. 2 shows the $n_{sf}$ versus δ dependence for homogeneous and heterogeneous clusters. As can be seen from Fig. 2, the change $n_{sf}$ with δ is described by the same curve in both cases. At the same time, a correlation of dependences of $n_{sf}$ and $V_{hcp}/V_{fcc}$ on δ is observed. Thus, in small clusters (δ ≤ 90 Å), for which the ratio $V_{hcp}/V_{fcc}$ is constant and equal to zero, the value of $n_{sf}$ is also constant. In this case, the number of intersecting defective planes is at its maximum, i.e. 4, which is characteristic of small fcc aggregations, as is shown in Ref. [9]. In the range of cluster size 90-150 Å, the rapid hcp phase increase was accompanied by a sharp decrease in the number of defective planes. Finally, when the linear characteristic cluster size reached 150 Å, the $n_{sf}$ value became zero while the $V_{hcp}/V_{fcc}$ value became constant.

The obtained experimental data allow us to propose a mechanism of nucleation and growth of the hcp phase in rare gas clusters, according to which the intersecting stacking faults present in the fcc lattice act as nuclei of an hcp structure. The increase in the percentage of the hcp phase with increasing cluster size is due to the fact that large clusters, which grew from large liquid drops, cool down more slowly and thus retain heat longer than small aggregations. As a result, the diffusion processes, which determine the growth of the hcp phase, are much more active in large clusters than in small ones. The more rapid increase in the hcp phase fraction in heterogeneous clusters is, presumably, also caused by the higher rate of diffusion processes in comparison with those in homogeneous clusters. Because in binary clusters, which are solid solutions of atoms of different size, microstrains can arise leading to an increase in the diffusion rate.

**FIGURE CAPTIONS**

Fig. 1. Ratio of volume of the hcp phase to volume of the fcc phase, $V_{hcp}/V_{fcc}$, as a function of size of single-component Ar clusters (□) and binary Ar-Kr clusters (●).

Fig. 2. Dependence of number $n_{sf}$ of the defective planes in the fcc matrix of Ar (□) and Ar-Kr (●) clusters on their size.

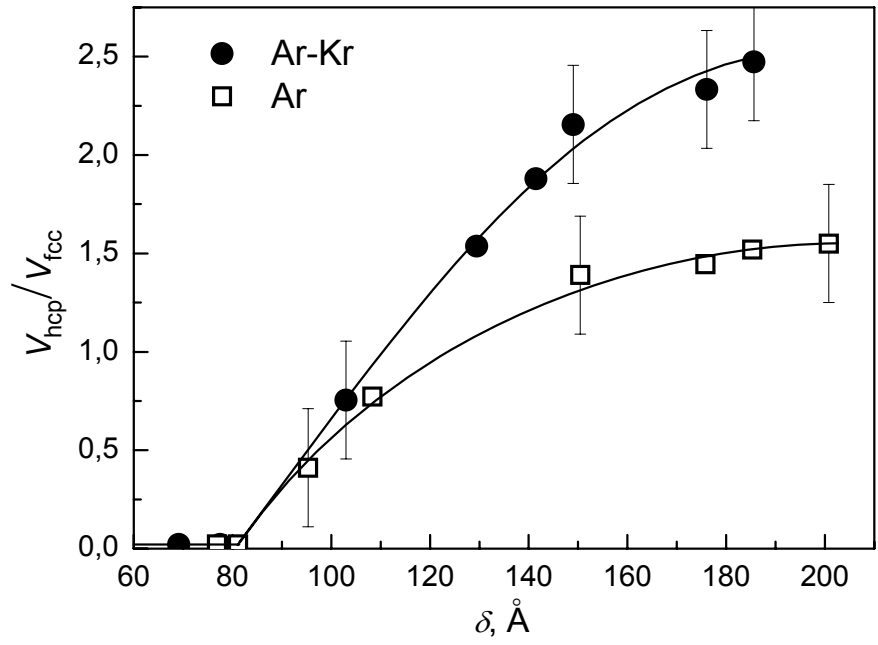

Figure 1

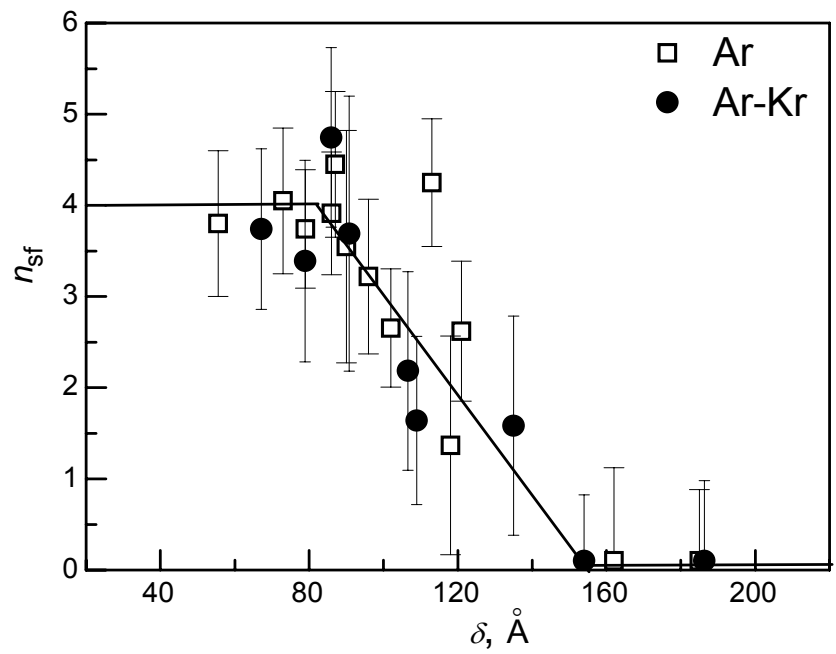

Figure 2